\documentclass[twocolumn,aps,prb,showpacs]{revtex4-1}
\usepackage[utf8]{inputenc}
\usepackage{color}
\usepackage{amsmath}
\usepackage{amssymb}
\usepackage{graphicx}
\usepackage[normalem]{ulem}
\usepackage[T1]{fontenc}
\usepackage[unicode=true, pdfusetitle, bookmarks=true, bookmarksnumbered=false, bookmarksopen=false, breaklinks=false, pdfborder={0 0 1}, backref=false, colorlinks=false]{hyperref}
\hypersetup{colorlinks,linkcolor=blue,citecolor=blue,urlcolor=blue}
\usepackage{float}
\usepackage{verbatim}
\usepackage[abs]{overpic}

\begin{document}
	\author{Arash Jafarizadeh}
	\affiliation{Instituto de F\'\i sica, Universidade Federal Fluminense, 24210-346 Niter\'oi RJ, Brazil}
	\author{M. A. Rajabpour}
	\affiliation{Instituto de F\'\i sica, Universidade Federal Fluminense, 24210-346 Niter\'oi RJ, Brazil}
	
	\title{Bipartite entanglement entropy of the excited states of the free fermions and the harmonic oscillators}	
	\begin{abstract}
		We study general entanglement properties of the excited states of the one dimensional translational 
		invariant free fermions and coupled harmonic oscillators. In particular, using the integrals of motion, 
		we prove that these Hamiltonians independent of the gap (mass) have infinite excited states that can be 
		described by conformal field theories with integer or half-integer central charges. In the case of 
		free fermions, we also show that because of the huge degeneracy in the spectrum, even a gapless 
		Hamiltonian can have excited states with an area-law. Finally, we study the universal average entanglement 
		entropy over eigenstates of the Hamiltonian of the free fermions introduced recently 
		in [L. Vidmar, et al., \href{https://link.aps.org/doi/10.1103/PhysRevLett.119.020601}{Phys. Rev. Lett. 119, 020601 (2017)}]. 
		In particular, using a duality relation, we map the average bipartite entanglement entropy  over all the exponential number 
		of eigenstates of a generic free fermion Hamiltonian  to the problem of the calculation of the average multi bipartite 
		entanglement entropy over a {\it{single}} eigenstate of the XX-chain. This relation can be useful for experimental measurement of 
		the universal average entanglement entropy. Part of our conclusions can be extended to the quantum spin 
		chains that are associated with the free fermions via Jordan-Wigner(JW) transformation.
		
	\end{abstract}
	
	\maketitle
	\graphicspath{{./figures/}}
	
\section{Introduction}
Bipartite entanglement entropy of quantum many-body systems exhibits a wide variety of interesting properties with a myriad 
of applications in high-energy and condensed matter physics. Some of the earliest studies were the calculation of
the entanglement entropy of the ground state (GS) of the coupled harmonic oscillators and the establishment of the area-law
in the same models in \cite{arXiv:1402.3589,PhysRevD.34.373,PhysRevLett.71.666}. In Ref \cite{HOLZHEY1994443}, the same
quantity was calculated for the ground state of the conformal field theories (CFT) and the famous logarithmic-law with
the central charge as the coefficient of the logarithm was derived. These studies were followed with many other interesting
results which paved the way to better understanding of the bipartite entanglement entropy of the ground state  of the
free fermions \cite{PhysRevB.64.064412,peschel2003calculation,PhysRevLett.96.100503}, coupled harmonic
oscillators \cite{PhysRevA.66.042327,osterloh2002scaling,PhysRevA.66.032110}, quantum spin chains \cite{PhysRevLett.90.227902,Jin2004,Hastings-2007}, CFTs \cite{calabrese2004entanglement} 
and topological systems\cite{PhysRevLett.96.110404,PhysRevLett.96.110405}. To review various applications of
the bipartite entanglement entropy of the ground state in many-body quantum systems and quantum field theories
see \cite{RevModPhys.80.517,Affleck2009,Refael2009,Peschel-2009,RevModPhys.82.277,RevModPhys.84.1655,laflorencie2016quantum,de2018genuine}
and  \cite{calabrese2009entanglement,castro2009bi,Casini-2009} respectively. Although the investigation of the bipartite entanglement
entropy of the ground state of quantum systems has a long history the same is not true for the excited states. The bipartite
entanglement entropy of the excited states in the quantum spin chains was first studied with exact methods in \cite{Alba2009}, 
see also \cite{Ares2014}. Then the entanglement entropy of the low-lying excited states in CFTs was calculated in \cite{Alcaraz2011,Berganza2012}. 
For recent numerical calculations regarding the entanglement entropy of the excited states in the quantum spin chains and free fermions
see \cite{Molter2014,Storms2014,Beugelin2015,Lai2015,Alba2015,Nandy2016,Vidmar2017,Vidmar-Rigol-2017,Riddell2018,Zhang2018,Liu2018,Vidmar2018,Hackl2019,miao2019,Grover2019,Srednicki2019}. For 
further results on the entanglement entropy of the low-lying excited states in CFTs see \cite{Taddia2013,Herwerth2015,Palmai2014,Taddia-2016,Ruggiero2017,Jiaju2019-1,Jiaju2019-2}. Recently, there 
have also been analytical calculations regarding the quantum entanglement content of the quasi-particle excitations in massive field theories and integrable chains\cite{Castro2018,CAlvaredo2018}. 

It is widely believed that one expects universality and quantum field theory for the ground state (and low-lying excited states) of the quantum many-body systems at and around quantum phase 
transition point. This has been one of the reasons that most of the studies were focused on the bipartite entanglement entropy of the ground states. Nevertheless, some typical 
behaviors (volume-law) have been already observed for the excited states too, see Refs \cite{Storms2014,Beugelin2015,Lai2015}. Some further analytical and numerical results were 
also obtained for the average of the entanglement entropy in \cite{Vidmar2017,Vidmar2018,Hackl2019} which support some sort of universality. In this paper, we would like to 
study the entanglement content of the excited states of the generic translational invariant one dimensional free fermions and coupled harmonic oscillators. The quantity of 
interest is the von Neumann entropy which is defined for a  system by partitioning it to $A$ and $\bar{A}$, where in this paper $A$ has $l$ contiguous sites, and $\bar{A}$ 
has $L-l$ sites where we will occasionally send $L$ to infinity. The von-Neumann entropy is 
\begin{equation}\label{von Neumann entropy}
	S_{vN}=-\text{tr}\rho_l\ln\rho_l,
\end{equation}
where $\rho_l$ is the reduced density matrix of the part $A$. Although in this article we will focus on the von Neumann entropy one can almost trivially extend the 
results to also R\'enyi entropies too. For the ground state of the gapped systems $S_{vN}$ saturates with the size of the subsystem\cite{Hastings-2007}, this is called 
an area-law. For the short-range critical one dimensional systems we have logarithmic-law \cite{HOLZHEY1994443,calabrese2004entanglement}, i.e.
\begin{eqnarray}\label{von Neumann-CFT}\
	S_{vN}=\frac{c}{3}\ln \Big{[}\frac{L}{\pi}\sin[\frac{\pi l}{L}]\Big{]}+\text{const},
\end{eqnarray} 
where $c$ is the central charge of the underlying CFT. When the $S_{vN}$ changes linearly with $l$ as happens typically for
the excited states, we call it a volume-law. In this paper, we will present some exact results that in some cases can be
even considered rigorous theorems. Most notably 
for one dimensional free fermions and coupled harmonic oscillators we prove the followings: $(1)$ All the Hamiltonians (independent of having a gap or not) 
have a lot of non-low lying excited states that can be described by CFT and have an arbitrary integer central charge. For free fermions
sometimes we can also have excited states with half-integer central charges. (2) The degenerate excited states depending on the chosen subspace basis 
can follow volume-law, logarithmic-law and sometimes even an area-law. (3) For free fermions (and corresponding spin chains)
there is a set of Hamiltonians that the ground state of every generic free fermion is one of the eigenstates of these Hamiltonians. In other
words, one can find the ground state of a generic Hamiltonian as one of the eigenstates of these Hamiltonians. Excited states with integer central charges have 
been realized before
in the context of the XX chain in \cite{Alba2009}, see also \cite{Ares2014}. However, our results extend those conclusions to
the generic translational invariant free fermions and coupled harmonic oscillators. In addition, our simple method not only explains
the existence of these kinds of excited states it also gives a natural way to make some statements regarding the average entanglement entropy studied in \cite{Vidmar2017,Vidmar2018,Hackl2019}.
In particular, combining the result of \cite{Vidmar2017} with  a duality relation introduced in \cite{Carrasco2017} we discover a method to calculate the average
bipartite entanglement entropy over all the eigenstates of a generic free fermion Hamiltonian using a single eigenstate of the XX-chain with just a simple hopping coupling.
Using this mapping instead of studying an exponential number of states one can just work with a {\it{single}} state which makes the previously considered unmeasurable quantity 
quite accessible for the experimental study.

The paper is organized as follows: In section II, we first define the Hamiltonian of generic translational invariant free fermions. Then 
for later arguments, we present some integrals of motion of these Hamiltonians. To clarify the argument, we then consider 
the XX chain and prove some statements regarding the entanglement content of this model. After that, we extend our arguments
to the general free fermions. Then we comment on the average entanglement over the excited states of free fermions, the role 
of the degeneracies, and how to measure specific averaging introduced in \cite{Vidmar2017}. In section III, we will define the Hamiltonian 
of generic coupled harmonic oscillators and then using the relevant integrals of motion; we will study the bipartite entanglement entropy of the excited states for these models.
Finally in the last section we will conclude the paper.
\section{Generic free fermions}
In this section we discuss some generic properties of the entanglement entropy of the excited states of the one dimensional free fermions.
 The Hamiltonian of a translational invariant (periodic) free fermions with time-reversal symmetry can be written as
\begin{eqnarray}\label{Hamiltonian}\
	H=\sum_{r=-R}^R\sum_{j\in\Lambda}a_rc_j^{\dagger}c_{j+r}+\frac{b_r}{2}(c_j^{\dagger}c_{j+r}^{\dagger}-c_jc_{j+r})+\text{const},\hspace{0.33cm}
\end{eqnarray} 
where $\Lambda=\{1,2,\cdots,L\}$. Using the Majorana operators $\gamma_j=c_j+c_j^{\dagger}$ and $\bar{\gamma}_j=i(c_j^{\dagger}-c_j)$ one can write $H=\frac{i}{2}\sum_{r=-R}^R\sum_{j\in\Lambda}t_r\bar{\gamma}_j\gamma_{j+r}$,
where $t_r=-(a_r+b_r)$ and $t_{-r}=-(a_r-b_r)$. It is very useful to put the coupling constants as the coefficients of the following holomorphic function $f(z)=\sum_rt_rz^r$, see\cite{Verresen2018}:
Then the Hamiltonian can be diagonalized by going to the Fourier space and then Bogoliubov transformation as follows:
\begin{eqnarray}\label{Hamiltonian-diagonalization}\
	H=\sum_k|f(k)|\eta_k^{\dagger}\eta_k+\text{const},
\end{eqnarray} 
where $\eta_k=\frac{1}{2}(1+\frac{f(k)}{|f(k)|})c_k^{\dagger}+\frac{1}{2}(1-\frac{f(k)}{|f(k)|})c_{-k}$ with $f(k):=f(e^{ik})$, where $k:=\frac{2\pi}{L}j$ with $j=1,2,...,L$. 
When the system is critical it is well-known that the number of zeros of the complex function $f(k)$ on the unit circle is twice the central charge of the underlying CFT.
This will be our guiding principle in most of the upcoming discussions.
The local 
mutually commuting integrals of motions can be written as follows:
\begin{eqnarray}\label{local-integral of motions m}\
	I_n^{+}&=&\sum_{k}\cos (nk)|f(k)|\eta_k^{\dagger}\eta_k,\hspace{0.6cm}n=0,1,...,\frac{L-1}{2},\\
	I_m^{-}&=&\sum_{k}\sin (mk)\eta_k^{\dagger}\eta_k,\hspace{1.4cm}m=1,...,\frac{L-1}{2}.\label{local-integral of motions p}
\end{eqnarray} 
The interesting and crucial fact is that the second set of integrals of motions in the real space can be written as
\begin{eqnarray}\label{local-integral of motions I minus}\
	I_m^{-}=\frac{i}{2}\sum_{j\in\Lambda}(c_j^{\dagger}c_{j+m}-c_{j+m}^{\dagger}c_{j}),
\end{eqnarray} 
which is independent of the parameters of the Hamiltonian.
The above considerations are also correct if one uses Jordan-Wigner transformation 
and find the quantum spin chain equivalent of the above Hamiltonians and integrals of motion, see Appendix A. For example,
all the periodic quantum spin chains that can be mapped to the free fermions commute with the following integrals of motion

%
%
%
\begin{eqnarray}\label{HXY1}
	I^-_{m}(XY)=\nonumber\hspace{8cm}\\
	\sum_{j\in\Lambda}\Big{[}\sigma_j^x\sigma_{j+1}^z...\sigma_{j+m-1}^z\sigma^y_{j+m}
	-\sigma_j^y\sigma_{j+1}^z...\sigma_{j+m-1}^z\sigma^x_{j+m}\Big{]}, \hspace{1cm}
\end{eqnarray}
which is independent of the parameters of the Hamiltonian.

\begin{figure}[t]
    \centering
    \includegraphics[width=0.45\textwidth]{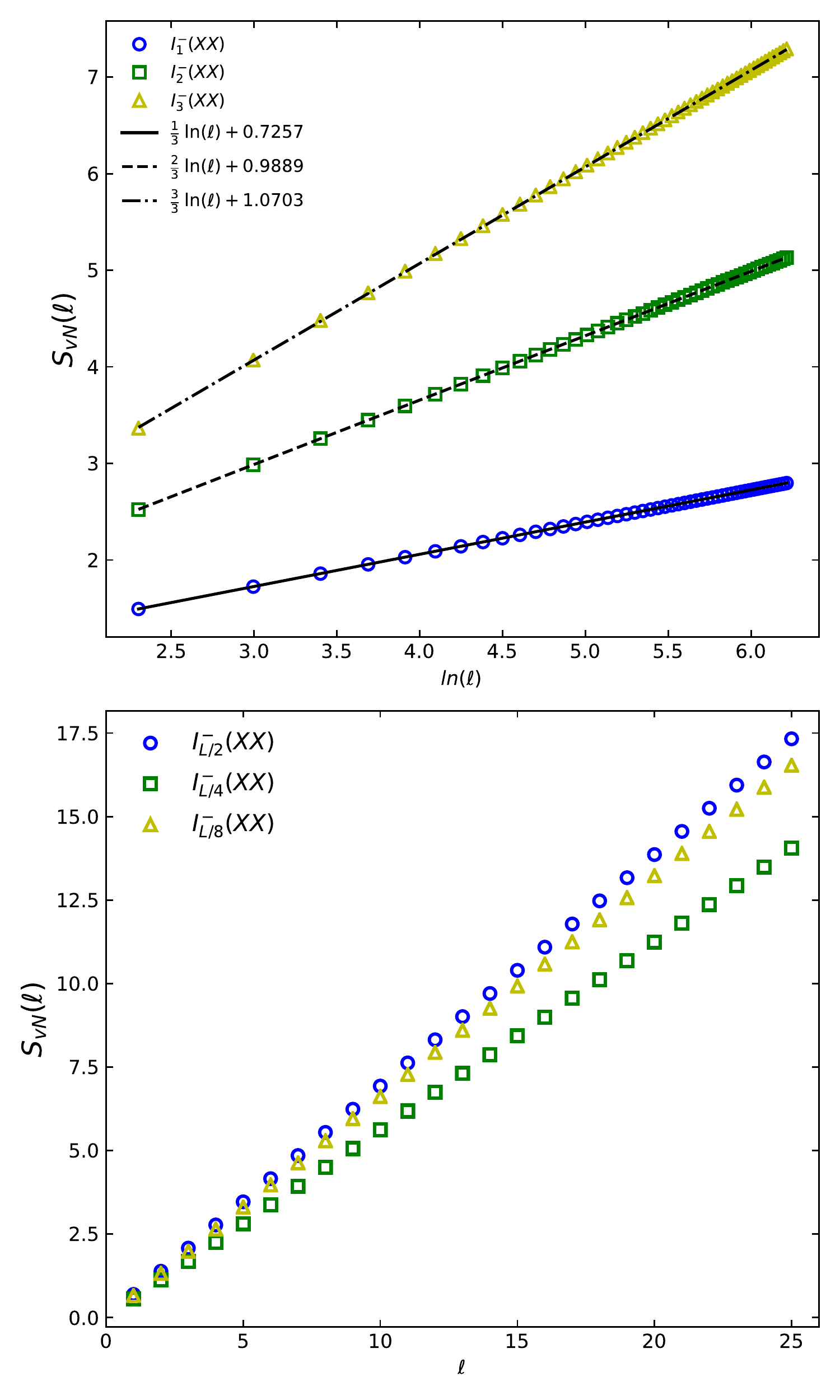}
	\caption{\footnotesize Top: bipartite von Neumann entropy with respect to the logarithm of the 
	size of the subsystem for the ground state of $I_m^{-}(XX)$ with $m=1,2,3$. 
	Bottom: bipartite von Neumann entropy with respect to the 
	size of the subsystem for the ground state of $I_m^{-}(XX)$ with $m=\frac{L}{8},\frac{L}{4},\frac{L}{2}$, and $L=200$}\label{fig:I minus}
\end{figure}

To introduce the main idea we first start with the XX-chain with the following Hamiltonian
\begin{eqnarray}\label{XX-hamiltonian}\
	H_{XX}=-\sum_{j\in\Lambda}(c_j^{\dagger}c_{j+1}+c_{j+1}^{\dagger}c_{j}).
\end{eqnarray} 
We are interested in the structure of particular excited states in the spectrum of the above Hamiltonian. The local commuting integrals of motions after a bit of rearrangement are 
\begin{eqnarray}\label{XX-integrals of motion: Fourier space1}\
	I_n^{+}(XX)&=&-2\sum_{k}\cos (nk)\eta_k^{\dagger}\eta_k,\hspace{0.2cm}n=0,1,...,\frac{L-1}{2},\hspace{0.33cm}\\
	\label{XX-integrals of motion: Fourier space2}\
	I_m^{-}(XX)&=&2\sum_{k}\sin (mk)\eta_k^{\dagger}\eta_k,\hspace{0.4cm}m=1,...,\frac{L-1}{2}.
	\end{eqnarray} 
In the real space $I_n^{-}(XX)$ is (\ref{local-integral of motions I minus}) and $I_n^{+}(XX)$ has the following form:
\begin{eqnarray}\label{XX-integrals of motion: real space}\
	I_n^{+}(XX)&=&-\sum_{j\in\Lambda}(c_j^{\dagger}c_{j+n}+c_{j+n}^{\dagger}c_{j}).
\end{eqnarray} 
The Hamiltonian and the integrals of motion share a common eigenbasis. That means, for example, the ground state of $I_n^{+}(XX)$ appears 
as the excited state of $H_{XX}$. We use this basic fact to prove our statements. Following \cite{Verresen2018}, and 
references therein, it is easy to see that for $I_n^{+}(XX)$ we have $f(z)=-(z^n+z^{-n})$ which have $2n$ zeros on the unit 
circle and so its ground state is critical and in the limit of large $L$ it can be described by a CFT with the central charge $c=n$. This 
can be also checked by calculating the entanglement entropy analytically (using the FH theorem  \cite{Alba2009}) and numerically
(using the Peschel method \cite{peschel2003calculation}, see  appendix B) by hiring the correlation matrix $C_{jk}=\langle c_j^{\dagger}c_k\rangle$. 
For the ground state of infinite size $I_n^{+}(XX)$
we have \footnote{We note that this sum can be simplified formally but for numerical calculations the above form is more useful\label{footnoteCm}}
\begin{eqnarray}\label{C matrix plus}\
	C_{jk}=\frac{1}{ \pi  (j-k)}\sum_{m=1}^n(-1)^{m+1} \sin \left(\frac{\pi  (2 m-1) (j-k)}{2 n}\right),\hspace{0.5cm}
\end{eqnarray} 
and $C_{jj}=\frac{1}{2}$. The above $\mathbf{C}$ matrix can be also found as the correlation matrix of one of the excited
states with energy zero of the Hamiltonian (\ref{XX-hamiltonian}). The correlation matrix of the  ground state of the infinite size $I_m^{-}(XX)$ 
can be also calculated easily and it is \footnotemark[1] 
\begin{eqnarray}\label{C matrix minus}\
	C_{jk}=\frac{-i}{2 \pi  (j-k)}\times\hspace{4cm}\nonumber\\
	\sum_{n=[-m/2]+1}^{[m/2]}\Big{(}e^{\frac{2in\pi}{m}(j-k)}-e^{i\frac{(2n-1)\pi}{m}(j-k)}\Big{)},
\end{eqnarray} 
and $C_{jj}=\frac{1}{2}$. The central charge of the underlying CFT is $c=m$. In the Figure \ref{fig:I minus} we show the logarithmic behavior
(and the corresponding coefficient) of the entanglement entropy of the ground state of $I_m^{-}(XX)$ 
for $m=1,2,3$\footnote{We note that although the two correlation matrices (\ref{C matrix plus}) and (\ref{C matrix minus}) are different 
they have the same set of eigenvalues\label{correlation matrices eigenvalues}}.
Based on the above arguments one can conclude that there are infinite conformal excited states with logarithmic scaling 
of the entanglement in the spectrum of $H_{XX}$ with the central charge $c=m$, where $m$ can be any integer number.
In the above, we looked to the finite $m$ when $L$ goes to infinity. However, it is clear that if $m$ is comparable to $L$, 
then one would expect for large $L$ probably a volume-law instead of a logarithmic behavior. This is indeed correct as it 
was shown in the Ref \cite{Ares2014} for $I_n^{+}(XX)$. The entanglement entropy of these excited states follows a volume-law with a subleading term which is logarithmic. 
In the Figure \ref{fig:I minus} we show the linear behavior of the entanglement entropy of the ground state of $I_m^{-}(XX)$ for $m=\frac{L}{8},\frac{L}{4},\frac{L}{2}$\footnotemark[2].
In 
the $L\to\infty$ there are an infinite number of this kind of energy excited states too. It is just enough to take $n=\alpha L$ where $0<\alpha<1$.
Note 
that in these cases the corresponding Hamiltonians $I_n^{+}(XX)$ are not local Hamiltonians. We note that there are also excited states
that follow an area-law. For example, the states $|11...1\rangle$ and $|00...0\rangle$ have energy zero and trivially follow an area-law.
In the case of $I_m^{-}$ as we will discuss soon this observation is more pronounced.
\begin{figure}[t]
    \centering
    \includegraphics[width=0.45\textwidth]{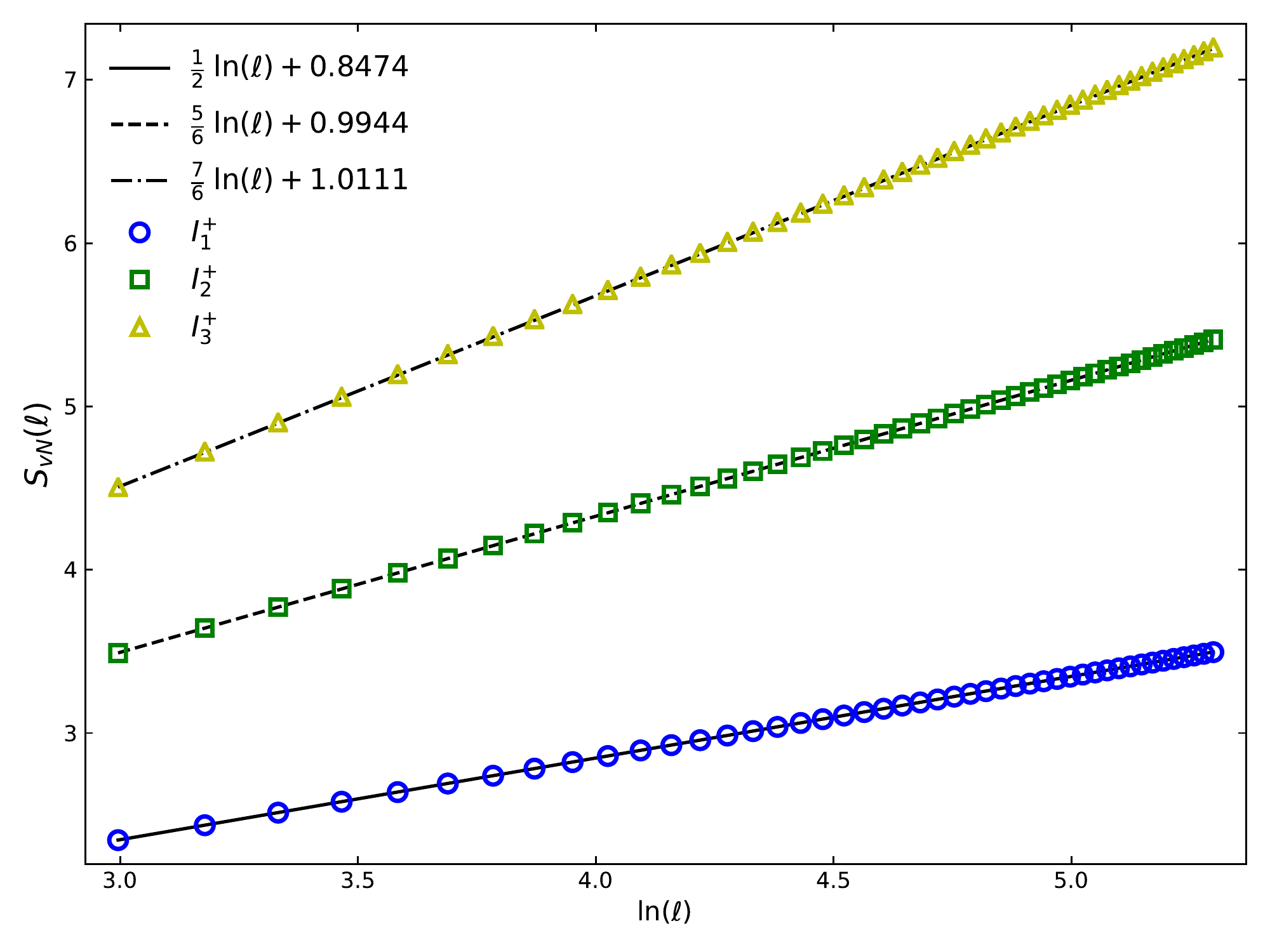}
	\caption{\footnotesize Bipartite von Neumann entropy with respect to the logarithm of the 
	size of the subsystem for the ground state of $I_m^{+}$ with $m=1,2,3$ produced out of the critical Ising parameters.}\label{fig:II:Ising}
\end{figure}

For generic free fermions, the arguments are similar: the Hamiltonian $H=I_0^{+}$ commutes with $I_m^{-}$ which means that 
the ground state of these integrals of motion should appear in the excited states of $H$. This means that even if the Hamiltonian 
is gapped (such as the gapped XY-chain) with the area-law property for the GS, there are still infinite excited states in the spectrum 
that are conformal invariant with the central charge $c=m$ and follow the logarithmic behavior. Of course, there are also an infinite number of excited states that follow the volume-law too. 
Interestingly, one can also argue that although the Hamiltonians $I_m^{-}$  have critical ground states, 
they also have an infinite number of excited states that follow an area-law. This simply because independent of 
the parameters in $H=I_0^{+}$ the Hamiltonians $I_m^{-}$ commute with it. Clearly, we have infinite possibilities to produce
Hamiltonians $I_n^{+}$ with gapped ground states that follow an area-law. If one starts with a 
critical Hamiltonian $H$ with half-integer central charge (for example an Ising critical chain) then it is easy to see that the 
Hamiltonians $I_m^{+}$ will have ground states with all the possible half integer numbers, i.e. $c=n+\frac{1}{2}$. That means, for example,
the Hamiltonian of the critical Ising chain has excited states with all the possible integer and half-integer numbers. In the Figure \ref{fig:II:Ising} we show
the logarithmic behavior (and the corresponding coefficients) of the entanglement entropy of the ground state of the $I_n^{+}$, associated to the 
critical transverse field Ising chain, for $n=1,2,3$.


\subsection{VHBR averaging and a proposal for measurement}

In the case of the periodic free fermions, we expect a lot of degeneracy in the excited states. As it 
was discussed in more detail in the Appendix C the number of independent energies over the size of
the Hilbert space decays exponentially with respect to the size of the system which indicates an enormous number of degeneracies.
\begin{figure}[t]
    \centering
    \includegraphics[width=0.483\textwidth]{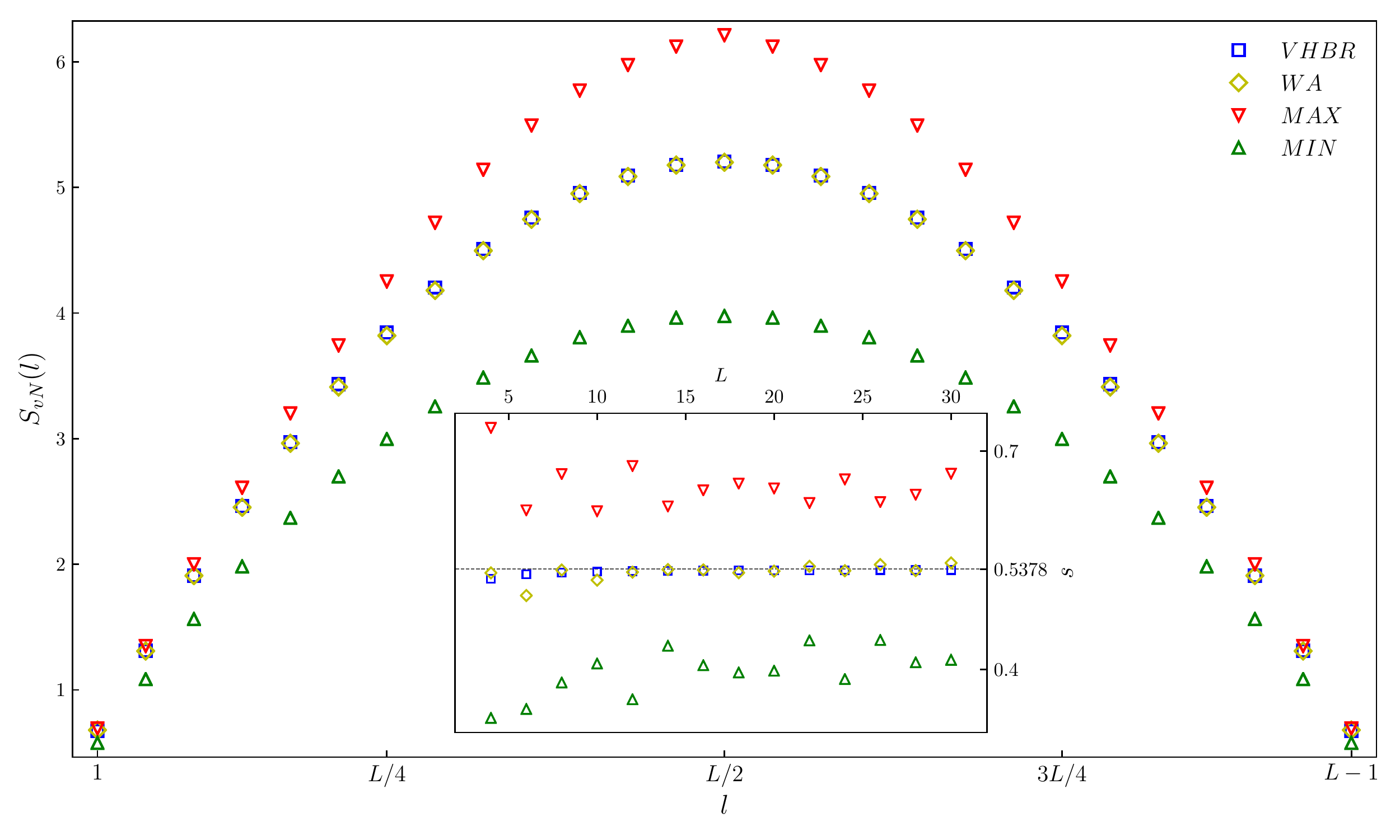}
	\caption{\footnotesize Various averagings of the entanglement entropy of the eigenstates of free fermions (XX-chain) in a periodic chain with $L = 28$ sites. Results 
	are plotted as a function of the linear subsystem size $l$. In the inset $s=\tfrac{S_{vN}(L/2)}{\tfrac{L}{2}\ln2}$ is plotted as a function of the system size 
	for $L\leq 30$. As one can see, VHBR (blue square) and WA (yellow diamond) averagings converge to the number $0.5378$ reported in \cite{Vidmar2017}. However, 
	the MAX (red triangle) and MIN (green triangle) do not converge to the same value.}\label{fig:diff Average}
\end{figure}
The way that one defines the excited state might be significant in getting area, logarithmic or volume law for the entanglement entropy. Recently in \cite{Vidmar2017,Vidmar2018} 
it was observed that if one takes all the eigenstates produced by the creation operators $\eta_k^{\dagger}$
and calculates the bipartite entanglement entropy of a connected region the result is universal and independent of the Hamiltonian. Here we would like to comment that the
result can be different if one takes into account the degeneracies. For this purpose in Figure \ref{fig:diff Average} we did the averaging for the XX chain 
in four different ways: The first one is the averaging proposed in \cite{Vidmar2017,Vidmar2018} (VHBR). Second (third) we identify the degenerate states
and pick the one with the minimum (maximum) entanglement entropy, then we do the averaging over the selected independent states. We call this
 MIN (MAX) averaging. Finally, in the 
last averaging we first identify the degenerate states and average over the entanglement entropy inside the subspace and then average over all
the averaged entanglement entropies (WA averaging), for more details see Appendix C. In Figure \ref{fig:diff Average}, one can see that the VHBR and 
the WA averagings converge to the same value but
the MIN and the MAX averagings converge to the different values. This clearly shows that the averaging done in the \cite{Vidmar2017,Vidmar2018} is special. 
Remarkably, the averaging and universality proposed in these papers have a good chance to be measured experimentally. This is because
of a recent duality proposed in \cite{Carrasco2017}. In this work, it was shown that there is a one to one correspondence between bipartite
entanglement entropy of the excited states (produced by the $\eta^{\dagger}_k$) of the XX-chain and the entanglement entropy of one eigenstate (the ground state for the case of $l=\frac{L}{2}$)
but with different partition. There are $2^L$ possible multi interval bipartitions for the eigenstate of the XX-chain which is identically equal to the
one interval bipartite entanglement entropy of $2^L$ excited states. Using this method it is possible to regenerate the universal figure proposed in \cite{Vidmar2017}, see Figure \ref{fig:diff Average}.
Consider the XX-chain Hamiltonian with the eigenstates $|K\rangle$, where $K=\{k_1,k_2,...,k_M\}\subset\{0,...,L-1\}$ are the excited modes. Then 
consider a subset of sites $A=\{x_1,x_2,...,x_M\}\subset\{0,...,L-1\}$. Based on \cite{Carrasco2017} if we consider $S(A;\psi)$ as the 
entanglement of the sites $A$ with respect to the rest for the pure state $|\psi\rangle$, then we have
\begin{equation}\label{eq:Sup partition duality}
S(A;K)=S(K;A),
\end{equation}
for any entropy functional. We now discuss a few examples of the above equality. For the half-filling to get the ground state, we need to fill all the negative modes which 
make half of the modes that are adjacent in the set $K$. This means the entanglement entropy of the ground state for the set $A$ is equal to the entanglement entropy of 
the half of the system for an eigenstate with the modes $A$ excited. Of course, if one averages over the entanglement entropy of all the possible sets $A$ one recover 
the average entanglement entropy VHBR for half of the system. Now consider we are interested in the averaging of VHBR for one site, in another word, $A=\{x_1\}$. This can 
be calculated by finding the average of the entanglement entropy of all the multi interval bipartitions of the state $K=\{k_1\}$ with just one mode excited. 
The rest of the VHBR graph can be produced similarly by calculating the average of the entanglement entropy of all the multi interval bipartitions of one eigenstate.

The above remarkable observation means that one can calculate the average entanglement
entropy by just calculating the entanglement entropy of all the multi interval bipartitions for a single eigenstate. The
entanglement entropy of different partitions in the spin chains has been already studied theoretically and experimentally in \cite{Elben2018,Brydges260}. The method
proposed and implemented in these works gives access to the entanglement entropy of all the $2^L$ different partitions for a particular
state. Extension of this technique to free fermions can make it an outstanding method to measure the universal average entanglement entropy proposed in \cite{Vidmar2017,Vidmar2018}. 
It is important to mention that because of the non-local nature of the JW transformation the average over the entanglement of all the 
possible bipartitions of the GS in the free fermions and the equivalent spin chains are not equal \cite{Igloi-Peschel-2009}.

\section{Coupled harmonic oscillators (CHO)}

\begin{figure} [t] 
\centering
\includegraphics[width=0.48\textwidth]{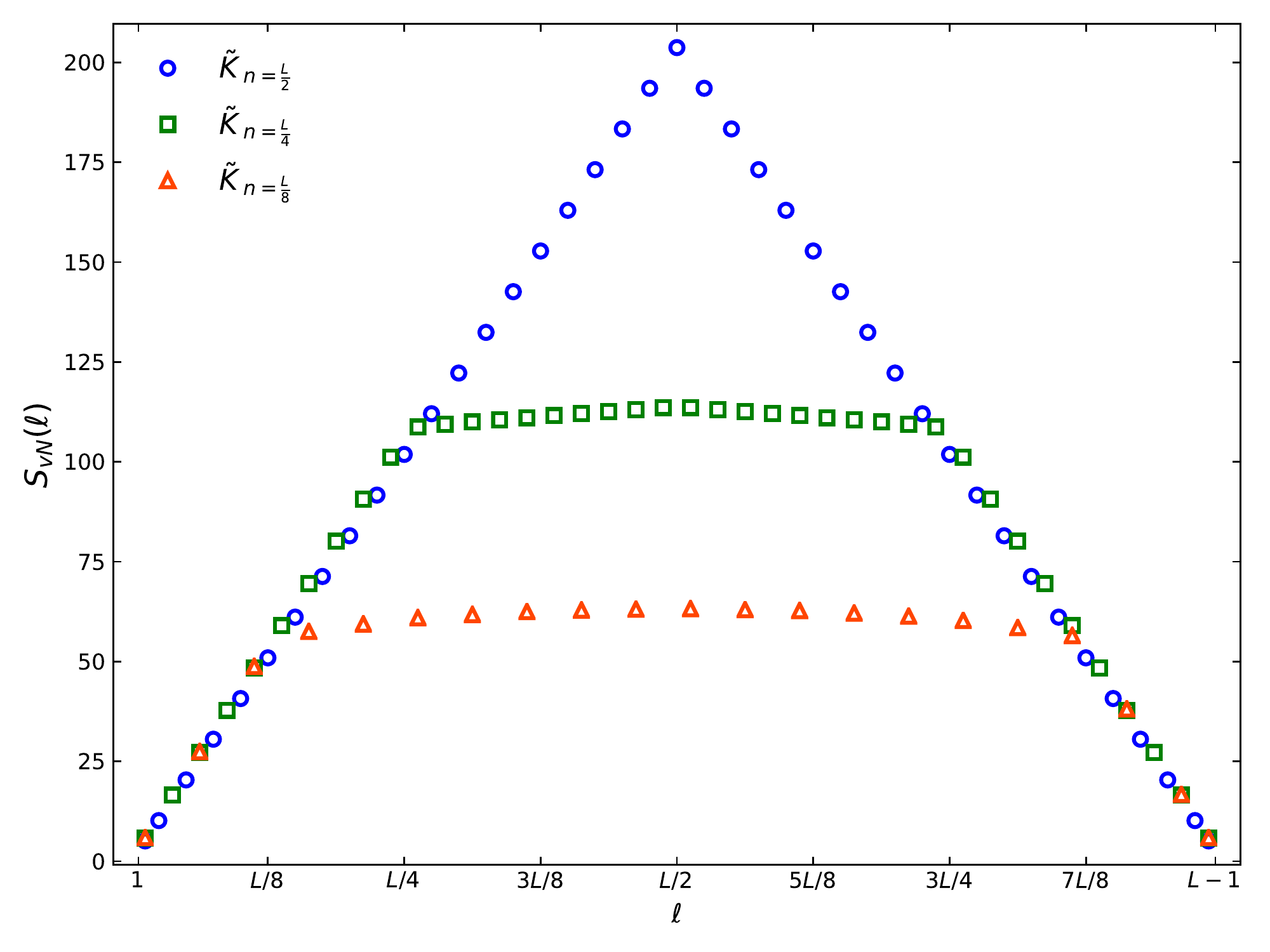}
\caption{\footnotesize Entanglement entropy of CHO with respect to the subsystem length ($\ell$) for $L=160$. We take $r\in\{2,4,6,\cdots,L-1\}$ (blue circles),  $r\in\{2,6,10,\cdots,L-1\}$ (green squares) and $r\in\{2,10,18,\cdots,L-1\}$ (red triangles). With this set of parameters, the volume law behavior is apparent.}  
\label{fig:Volum law HO}
\end{figure}

In this section we generalize the ideas of the previous section to coupled harmonic oscillators. 
 The Hamiltonian of a periodic CHO can be written as
\begin{eqnarray}\label{Hamiltonian CHO}\
H(\mathbf{K},\mathbf{P})=\frac{1}{2}\sum_{i,j=1}^L(\hat{\pi}_iP_{ij}\hat{\pi}_j+\hat{\phi}_iK_{ij}\hat{\phi}_j),
\end{eqnarray}
where $\mathbf{P}$ and $\mathbf{K}$ are symmetric positive definite matrices. For translational invariant cases, these matrices 
are circulant matrices. The ground state of the above Hamiltonian can be written as
\begin{eqnarray}\label{ground state CHO}\
\Psi_g(\{\phi_i\})=\Big{(} \det\frac{K^{1/2}}{\pi}\Big{)}^{1/4}e^{-\frac{1}{4}\langle\phi|K^{1/2}|\phi\rangle}.
\end{eqnarray}
The ground state does not depend on the matrix $\mathbf{P}$ which clearly also means that the entanglement entropy
of the ground state is also independent of the matrix $\mathbf{P}$.
To find  a set of integrals of motion for the above Hamiltonian we consider the operator, $\tilde{H}=H(\tilde{\mathbf{K}},\tilde{\mathbf{P}})$.
This operator commutes with the Hamiltonian as far as 
\begin{eqnarray}\label{Key condition CHO}
\mathbf{P}.\tilde{\mathbf{K}}-\tilde{\mathbf{P}}.\mathbf{K}=0.
\end{eqnarray}
Without loosing generality now consider $\mathbf{P}=\mathbf{I}$ then we have $\tilde{\mathbf{K}}=\tilde{\mathbf{P}}.\mathbf{K}$. Since the
$\tilde{\mathbf{P}}$ can be chosen arbitrarily one can have very generic circulant matrix $\tilde{\mathbf{K}}$. 
The ground state of $\tilde{H}$ which is the excited state of $H$ can now have very generic form with quite general entanglement content. 
As the most natural example we first
consider discrete massive Klein-Gordon Hamiltonian with the following $\mathbf{K}$ matrix
\begin{eqnarray}\label{discrete KG K matrix}\
    K_{nm}=\frac{1}{L}\sum_{j=0}^{L-1} e^{i\frac{2\pi (m-n)j}{L}}\Big{[}2(1-\cos(\frac{2\pi j}{L})+m^2\Big{]}.
\end{eqnarray}
The above matrix is just a massive discrete Laplacian on a circle. In the limit of large $L$ when $m=0$ the underlying conformal field theory has the central charge
$c=1$. Consequently the entanglement entropy grows logarithmically in this case. However, for the massive case we have an area-law \cite{PhysRevD.34.373,PhysRevLett.71.666,Plenio2005,Cramar2006}. 
These results can be easily checked numerically using the method described in \cite{Plenio2005,Cramar2006}, see Appendix B.
Now consider the Hamiltonian $\tilde{H}$ with the following $\tilde{\mathbf{P}}$ and $\tilde{\mathbf{K}}$ matrices
        \begin{eqnarray}\label{P tilde matrix}
            \tilde{P}_{nm}=\frac{1}{L}\sum_{j=0}^{L-1} e^{i\frac{2\pi (m-n)j}{L}}\prod_{r}(2-2\cos(\frac{2\pi j}{L}+\beta_r))^{\alpha_r},\hspace{0.5cm}\\
            \label{K tilde matrix}
            \tilde{K}_{nm}=\frac{1}{L}\sum_{j=0}^{L-1} e^{i\frac{2\pi (m-n)j}{L}}\prod_{r}(2-2\cos(\frac{2\pi j}{L}+\beta_r))^{\alpha_r}\hspace{0.5cm}\nonumber\\\times2(1-\cos(\frac{2\pi j}{L})+\frac{m^2}{2}),\hspace{3.3cm}
        \end{eqnarray}
where $\beta_r$'s are different real numbers \footnote{Similar couplings for harmonic oscillators have been also discussed in \cite{Unany2005}}. For example, we 
can take $\beta_r=2\pi\frac{r}{L}$ with $r\in\{1,2,...,L-1\}$. Following \cite{Nezhad2014}, consider  $e^{i\beta_r}$'s appear in complex conjugate form and we have $2n$ of them which $n$ is 
not proportional 
to the size of the system and $\alpha_r=1$. Then we expect to have a free field theory $\tilde{H}$ with a  ground state that have the central
charge $n+1$, see \cite{Nezhad2014}. When $n$ is proportional to the size of the system we expect
to get a volume-law as we had in the case of the free fermions. Numerical calculations presented in the Figures \ref{fig:Volum law HO} confirm the above expectation.
We note that since
the $\tilde{\mathbf{K}}$ matrix always has a singular part (with this method) it is not possible to generate an excited state which follows the area-law.
When $m\neq 0$ the ground state and the low-laying states follow an area-law. However, if we choose the $\tilde{\mathbf{P}}$ as (\ref{P tilde matrix}) 
then one can generate a $\tilde{\mathbf{K}}$ matrix which is singular and follows the logarithmic-law or the volume-law. In other words, as the case 
of free fermions even for massive cases we have a lot of  excited states that can be described by CFT.

\section{Conclusions}

 We showed that independent of the gap a generic translational invariant free-fermion Hamiltonian in one dimension 
has infinite eigenstates that follow logarithmic-law of entanglement and can be described by CFT. We argued that because of the huge degeneracy, 
even a Hamiltonian with a critical GS can in principle have a lot of excited states with an area-law behavior. Similar conclusions 
are valid also for the excited states of the corresponding spin chains. We also proposed a method to measure the recently proposed universal 
average entanglement entropy over all the exponential number of eigenstates of a generic free fermion Hamiltonian by averaging over multi interval bipartite EE of just a {\it{single}} 
eigenstate of the XX-chain. We finally extended our discussion to the generic HOs in one dimension. It will be interesting to 
explore in more detail the averaging over the multi interval bipartite EE of the eigenstates of the generic free fermions and also spin chains.

\vspace*{0.5cm}



%

	
\begin{acknowledgments}		
\textbf{Acknowledgments:} We thank P. Calabrese and B. Vermersch for useful discussions and M. Sarandy for support through the computational resources.
M. A. R. thanks ICTP for the financial support and hospitality during the final stages of the work. This work was supported in parts by CNPq. 
\end{acknowledgments}

\setcounter{equation}{0}
\renewcommand{\theequation}{A\arabic{equation}}



\section*{Appendices}
In the next three Appendices for the sake of completeness, we summarize first the spin chain version of the fermionic integrals of motion that we introduced in the main text. Then we introduce the
method of correlations that can be used to calculate the entanglement entropy in the free fermions and the harmonic oscillators. In the third subsection, we comment on the degeneracies 
in the XX-chain and various averagings on the entanglement entropy of the excited states. 

\subsection*{Appendix A: Integrals of motion: spin chains}\label{app1}
This section serves to pay more attention to the spin version of the integrals of motion and the fermionic Hamiltonian defined in the main text.
In most of the cases, the integrals of motion do not have a simple form. However, one can still claim that the Hamiltonian $I_{m}^-$ which is parameter
independent in the spin version, commutes with all the possible Hamiltonians and integrals of motion and so it provides bases
which in those bases the average of entanglement entropy is equal for the  eigenstates of $I_{m}^-$ and the Hamiltonian. 
We begin by writing the spin form of fermionic operators using the Jordan-Wigner transformation 
\begin{equation}\label{eq:Sup JWT}
	c_j=\prod_{l<j}\sigma_l^z\sigma_j^-.
\end{equation}
where $\sigma_i^{\pm}=\frac{\sigma_i^x\pm\sigma_i^y}{2}$. Substituting the above in the fermionic Hamiltonian we get
\small
\begin{equation}\label{eq:Sup H general}
	\begin{aligned}
	H=&\sum_{r>0}\sum_{j=1}^{L-1}\Big[\frac{-a_r-b_r}{2}\sigma^x_{j}\sigma^z_{j+1}\cdots\sigma^z_{j+r-1}\sigma^x_{j+r}\\
	&+\frac{-a_r+b_r}{2}\sigma^y_{j}\sigma^z_{j+1}\cdots\sigma^z_{j+r-1}\sigma^y_{j+r}\big]-\sum_{j=1}^{L}\frac{a_0}{2}\big(\sigma_j^z-1\big)\\
	&-\hat{\mathcal{N}}\big(\frac{-a_r-b_r}{2}\sigma^x_{L}\sigma^z_{1}\cdots\sigma^z_{r-1}\sigma^x_{r}\\
	&+\frac{-a_r+b_r}{2}\sigma^y_{L}\sigma^z_{1}\cdots\sigma^z_{r-1}\sigma^y_{r}\big),
	\end{aligned}
\end{equation}\normalsize
where $\hat{\mathcal{N}}=\prod_{l=1}^{L}\sigma_l^z$, is the parity of the spins down,
i.e. the parity of the number of fermions. Because of the string of $\sigma^z$, the JW transformation is non-local. 
The non-locality of the JW transformation affects the boundary conditions through appearance of the operator $\hat{\mathcal{N}}$. 
The eigenvalues: $\mathcal{N}=+1\:\text{and}\:-1$ correspond to anti-periodic and periodic boundary conditions respectively. 
The operator $\hat{\mathcal{N}}$ should appear every time we go from spin format to fermion format or vice versa. For instance, in the periodic boundary condition, one gets:\small
\begin{equation}
\begin{aligned}
	H(\scalebox{0.7}{$\mathcal{N}=-1$})&=\sum_{r>0}\sum_{j=1}^{L}\Big[\frac{-a_r-b_r}{2}\sigma^x_{j}\sigma^z_{j+1}\cdots\sigma^z_{j+r-1}\sigma^x_{j+r}\\
	&+\frac{-a_r+b_r}{2}\sigma^y_{j}\sigma^z_{j+1}\cdots\sigma^z_{j+r-1}\sigma^y_{j+r}\big]\\
	&-\sum_{j=1}^{L}\frac{a_0}{2}\big(\sigma_j^z-1\big).
\end{aligned}
\end{equation}\normalsize
The above Hamiltonian when $a_r$ and $b_r$ are nonzero just for $r=0,1$ becomes: 
\begin{equation}\label{eq:Sup XYHam}
	\begin{aligned}
	H_{\scalebox{0.5}{$r=0,\pm1$}}&(\scalebox{0.7}{$\mathcal{N}=-1$})=-\frac{a_1}{2}\sum_{j=1}^{L}\big[(1+\frac{b_1}{a_1})\sigma^x_{j}\sigma^x_{j+1}\\
	&+(1-\frac{b_1}{a_1})\sigma^y_{j}\sigma^y_{j+1}\big] -\frac{a_0}{2}\sum_{j=1}^{L}\sigma_j^z+\frac{La_0}{2},
	\end{aligned}
\end{equation}
which is the Hamiltonian of the anisotropic periodic XY spin chain with $-a_1\mp b_1=\frac{J(1\pm\gamma)}{2}$ and $a_0=h$.\\

For Integrals of motion $I_n^+$ and $I_m^-$, we can find the spin form likewise. Using the inverse Fourier transformation, we write them in the $c$-fermion form. 
The $c$-fermion form of $I^-_m$ is already written in the main text; therefore, the spin form of this operator is
\small
\begin{equation}\label{eq:XXX}
\begin{aligned}
I_{m}^- =& -\frac{1}{4}\sum_{j=1}^{L-1}\big[\sigma^x_j\prod_{l=j+1}^{j+m-1}\sigma^z_{l}\sigma^y_{j+m}-\sigma^y_j\prod_{l=j+1}^{j+m-1}\sigma^z_{l}\sigma^x_{j+m}\big]\\
&+\frac{\hat{\mathcal{N}}}{4}\big[\sigma^x_L\prod_{l<m}\sigma^z_{l}\sigma^y_{m}-\sigma^y_L\prod_{l<m}\sigma^z_{l}\sigma^x_{m}\big].
\end{aligned}\end{equation}\normalsize
Putting $\mathcal{N}=-1$, the boundary term absorbs in the sum and $I^-_m$ reads as:
\begin{eqnarray}\label{eq:Sup -Ic}
    I_{m}^-(\mathcal{N}=-1)=\hspace{5.6cm}\\\nonumber -\frac{1}{4}\sum_{j=1}^{L}\big[\sigma^x_j\prod_{l=j+1}^{j+m-1}\sigma^z_{l}\sigma^y_{j+m}-\sigma^y_j\prod_{l=j+1}^{j+m-1}\sigma^z_{l}\sigma^x_{j+m}\big].
\end{eqnarray}
For $I^+_n$, the fermionic form 
would be 
\begin{equation}\label{eq:Sup +I c}
	\begin{aligned}
	I_n^+=&\sum_{j\in\Lambda}\sum_{r=-R}^{R}\Big(\frac{a_r}{2}\big[c^\dagger_{j}c_{j+r+n}+c^\dagger_{j+n}c_{j+r}\big]\\
	&+\frac{b_r}{4}\big[c^\dagger_{j}c^\dagger_{j+r+n}+c^\dagger_{j+n}c^\dagger_{j+r} -c_{j}c_{j+n+r}\\
	&-c_{j+n}c_{j+r}\big]\Big).
	\end{aligned}
\end{equation}
In the XY spin chain case this integral of motion can be written as
\begin{equation}
	\begin{aligned}
	I^+_{\scalebox{0.5}{$r=0,\pm1$}}=&\sum_{j\in\Lambda}\Big(\frac{J}{4}\big[c^\dagger_{j}c_{j+n+1}+c^\dagger_{j+n}c_{j+1}+c^\dagger_{j+1}c_{j+n}\\
	&+c^\dagger_{j+n+1}c_{j}\big]+\frac{h}{2}\big[c^\dagger_{j}c_{j+n}+c^\dagger_{j+n}c_{j}\big]\\
    &\;+\frac{J\gamma}{4}\big[c^\dagger_{j}c^\dagger_{j+n+1}+c^\dagger_{j+n}c^\dagger_{j+1} -c_{j}c_{j+n+1}\\
    &-c_{j+n}c_{j+1}\big]\Big),
    \end{aligned}
\end{equation}
where we have used the fact that $a_r=a_{-r}$ and $b_r=-b_{-r}$. $I^+_n$ commutes with both the Hamiltonian \eqref{eq:Sup XYHam} and $I_m^-$. 
To write the above expression in the spin form, we need to separate two possible cases: $n>1$ and $n=1$. Consequently, using the 
transformation \eqref{eq:Sup JWT}, the spin form of integrals of motion is:
\begin{widetext}

\begin{align}\label{eq:Sup +I spin1}
I_{\scalebox{0.5}{$r=0,\pm1$}}^+ =& -J\sum_{j=1}^{L-n-1}\big[\tfrac{(1+\gamma)}{8}\sigma_j^x\sigma_{j+1}^z\cdots\sigma_{j+n}^z\sigma_{j+n+1}^x+\tfrac{(1-\gamma)}{8}\sigma_j^y\sigma_{j+1}^z\cdots\sigma_{j+n}^z\sigma_{j+n+1}^y\big]\nonumber\\
&-J\sum_{j=1}^{L-n+1}\big[\tfrac{(1-\gamma)}{8}\sigma_{j}^x\sigma_{j+1}^z\cdots\sigma_{j+n-2}^z\sigma_{j+n-1}^x+\tfrac{(1+\gamma)}{8}\sigma_{j}^y\sigma_{j+1}^z\cdots\sigma_{j+n-2}^z\sigma_{j+n-1}^y\big]\nonumber\\
&-\frac{h}{4}\sum_{j=1}^{L-n}\big[\sigma_{j}^x\sigma_{j+1}^z\cdots\sigma_{j+n-1}^z\sigma_{j+n}^x+\sigma_{j}^y\sigma_{j+1}^z\cdots\sigma_{j+n-1}^z\sigma_{j+n}^y\big]\hspace{5cm} n>1\\
&+J\hat{\mathcal{N}}\sum_{j=L-n}^{L}\big[\tfrac{(1+\gamma)}{8}\sigma_j^x\sigma_{j+1}^z\cdots\sigma_{j+n}^z\sigma_{j+n+1}^x+\tfrac{(1-\gamma)}{8}\sigma_j^y\sigma_{j+1}^z\cdots\sigma_{j+n}^z\sigma_{j+n+1}^y\big]\nonumber\\
&+J\hat{\mathcal{N}}\sum_{j=L-n+2}^{L}\big[\tfrac{(1-\gamma)}{8}\sigma_{j+1}^x\sigma_{j+2}^z\cdots\sigma_{j+n-1}^z\sigma_{j+n}^x+\tfrac{(1+\gamma)}{8}\sigma_{j+1}^y\sigma_{j+2}^z\cdots\sigma_{j+n-1}^z\sigma_{j+n}^y\big]\nonumber\\
&+\frac{h\hat{\mathcal{N}}}{4}\sum_{j=L-n+1}^{L}\big[\sigma_{j}^x\sigma_{j+1}^z\cdots\sigma_{j+n-1}^z\sigma_{j+n}^x+\sigma_{j}^y\sigma_{j+1}^z\cdots\sigma_{j+n-1}^z\sigma_{j+n}^y\big],\nonumber\\\nonumber
\end{align}
\begin{align}
I_{\scalebox{0.5}{$r=0,\pm1$}}^+ =& -J\sum_{j=1}^{L-2}\big[\tfrac{1+\gamma}{8}\sigma_j^x\sigma_{j+1}^z\sigma_{j+2}^x+\tfrac{1-\gamma}{8}\sigma_j^y\sigma_{j+1}^z\sigma_{j+2}^y\big]-\frac{h}{4}\sum_{j=1}^{L-1}\big[\sigma_j^x\sigma_{j+1}^x+\sigma_j^y\sigma_{j+1}^y\big]+\frac{J}{4}\sum_{j\in\Lambda}\sigma_{j}^z\nonumber\\
&+\hat{\mathcal{N}}J\big[\tfrac{1+\gamma}{8}\sigma_{L-1}^x\sigma_{L}^z\sigma_{1}^x+\tfrac{1-\gamma}{8}\sigma_{L-1}^y\sigma_{L}^z\sigma_{1}^y+\tfrac{1+\gamma}{8}\sigma_L^x\sigma_{1}^z\sigma_{2}^x+\tfrac{1-\gamma}{8}\sigma_L^y\sigma_{1}^z\sigma_{2}^y\big]\\
&+\frac{\hat{\mathcal{N}}h}{4}\big[\sigma_L^x\sigma_{1}^x+\sigma_L^y\sigma_{1}^y\big],\nonumber\hspace{10cm} n=1.
\end{align}
\end{widetext}
These quantities commute with each other and the Hamiltonian.
Note that these operators commute with each other after fixing the $\mathcal{N}$ too. For example, with $R=1$, we have:
\begin{eqnarray}
    \big[H_{\scalebox{0.5}{$r=0,\pm1$}}(\mathcal{N}=-1),I^+_{\scalebox{0.5}{$r=0,\pm1$}}(\mathcal{N}=-1)\big]=0, \\
    \big[ H{\scalebox{0.5}{$r=0,\pm1$}}(\mathcal{N}=-1),I^-_{\scalebox{0.5}{$r=0,\pm1$}}(\mathcal{N}=-1)\big]=0,\\
    \big[ I^-_{\scalebox{0.5}{$r=0,\pm1$}}(\mathcal{N}=-1),I^+_{\scalebox{0.5}{$r=0,\pm1$}}(\mathcal{N}=-1)\big]=0.
\end{eqnarray}
In a general case of $R$, we write for $I^+_n$:
\begin{align}\label{eq:Sup I+ spin general}
    I_{n}^+& = \sum_{r=-R}^{R}\Bigg[\sum_{j=1}^{L-n-r}\big[\tfrac{(a_r+b_r)}{4}\sigma_j^x\sigma_{j+1}^z\cdots\sigma_{j+n+r-1}^z\sigma_{j+n+r}^x\nonumber\\\nonumber
    &+\tfrac{(a_r-b_r)}{4}\sigma_j^y\sigma_{j+1}^z\cdots\sigma_{j+n+r-1}^z\sigma_{j+n+r}^y\big]\\\nonumber
    &-\hat{\mathcal{N}}\sum_{j=L-n-r+1}^{L}\big[\tfrac{(a_r+b_r)}{4}\sigma_j^x\sigma_{j+1}^z\cdots\sigma_{j+n+r-1}^z\sigma_{j+n+r}^x\\
    &+\tfrac{(a_r-b_r)}{4}\sigma_j^y\sigma_{j+1}^z\cdots\sigma_{j+n+r-1}^z\sigma_{j+n+r}^y\big]\Bigg],
\end{align}
for the case which $n>R$. In the case of $n\leq R$ one gets
\begin{widetext}
\begin{align}\label{eq:sup I+ spin gene}
    I_{n}^+ =&\sum_{\substack{r=0\\ r\neq n}}^{R} \Bigg[\sum_{j=1}^{L-n-r}\big[\tfrac{a_r+b_r}{4}\sigma_j^x\sigma_{j+1}^z\cdots\sigma_{j+n+r-1}^z\sigma_{j+n+r}^x+\tfrac{a_r-b_r}{4}\sigma_j^y\sigma_{j+1}^z\cdots\sigma_{j+n+r-1}^z\sigma_{j+n+r}^y\big]\nonumber\\
    &+\sum_{j=1}^{L+n-r}\big[\tfrac{a_r+b_r}{4}\sigma_{j+n}^x\sigma_{j+n+1}^z\cdots\sigma_{j+r-1}^z\sigma_{j+r}^x+\tfrac{a_r-b_r}{4}\sigma_{j+n}^y\sigma_{j+n+1}^z\cdots\sigma_{j+r-1}^z\sigma_{j+r}^y\big]\\
    &-\hat{\mathcal{N}}\sum_{j=L-n-r+1}^{L}\big[\tfrac{a_r+b_r}{4}\sigma_j^x\sigma_{j+1}^z\cdots\sigma_{j+n+r-1}^z\sigma_{j+n+r}^x+\tfrac{a_r-b_r}{4}\sigma_j^y\sigma_{j+1}^z\cdots\sigma_{j+n+r-1}^z\sigma_{j+n+r}^y\big]\nonumber\\
    &-\hat{\mathcal{N}}\sum_{j=L+n-r+1}^{L}\big[\tfrac{a_r+b_r}{4}\sigma_{j+n}^x\sigma_{j+n+1}^z\cdots\sigma_{j+r-1}^z\sigma_{j+r}^x+\tfrac{a_r-b_r}{4}\sigma_{j+n}^y\sigma_{j+n+1}^z\cdots\sigma_{j+r-1}^z\sigma_{j+r}^y\big]\Bigg]+\frac{a_{n}}{2}\sum_{j\in\Lambda}\sigma_{j}^z.\nonumber
\end{align}
\end{widetext}

There is also the possibility to write the spin form of our quantities of interest in a way that operator $\hat{\mathcal{N}}$ does not appear explicitly. As it can be seen, whenever a product of two fermionic operators at sites $i$ and $j$ is written in the Pauli spin operators,
there is a string of $\sigma^z$ between these two sites. As in equations \eqref{eq:sup I+ spin gene} and \eqref{eq:Sup I+ spin general}, we placed the string of spin-$z$ operator in the smallest path between our lattice sites. However, In the boundary terms, this smallest path is passing through the boundary. Therefore, we face terms like $\cdots\sigma^z_{L}\sigma^z_{1}\cdots$ accompanied with operator $\hat{\mathcal{N}}$. If before the JW-transformation, we rearrange our fermionic operators $c_i^{(\dagger)}$ and $c_j^{(\dagger)}$ in a way that the operator with smaller index comes on the left of the operator with a bigger index. In particular, one can write:\small
\begin{equation}
    c^\dagger_{L-n}c_{m}+c^\dagger_{m}c_{L-n}\;\longrightarrow\;\big(\sigma^x_{m}\sigma^x_{L-n}+\sigma^y_{m}\sigma^y_{L-n}\big)\prod_{j=m+1}^{L-n-1}\sigma^z_{j},
\end{equation}\normalsize
where $n,m<\frac{L}{2}$. This way of writing the integrals of motion eliminates the operator $\hat{\mathcal{N}}$ and both of these methods are equivalent. For example for $I^+_n$ with $n>R$ can be written as\small
\begin{align}
    I_{n}^+ =& \sum_{r=0}^{R}\sum_{1\leq i< j\leq L} \Big[\big(\tfrac{(a_r+b_r)}{4}\sigma_i^x\sigma_{j}^x+\tfrac{(a_r-b_r)}{4}\sigma_i^y\sigma_{j}^y\big)\prod_{k=j+1}^{l-1}\sigma_{k}^z\Big],
\end{align}\normalsize
which $j=i\pm r+n$, and since $n>r$ for all $r$ then $j>i$ for all r.

\setcounter{equation}{0}
\renewcommand{\theequation}{B\arabic{equation}}

\subsection*{Appendix B: Entanglement entropy in the  free fermions and the harmonic oscillators: the method of correlations}\label{app2}

For the free fermions/bosons one can use the matrix of correlations to calculate the entanglement entropy for relatively large sizes. 
In this subsection we provide the well-known exact formulas that can be found in the reviews\cite{Peschel-2009,Casini-2009}.
The entanglement entropy for the eigenstates of the free fermions (XX-chain) can be found using the following formula:
\begin{eqnarray}\label{Peschel method}\
	S_{vN}=-\sum_j\lambda_j\ln\lambda_j+(1-\lambda_j)\ln(1-\lambda_j),
\end{eqnarray} 
where $\lambda_j$'s are the eigenvalues of the correlation matrix $\mathbf{C}_A$ restricted to the subsystem $A$ with the elements $C_{ij}=\langle c_i^{\dagger}c_j\rangle$. 

In the case of Harmonic oscillators the von Neumann entanglement entropy has the following form:
\begin{equation*}\label{EE CHO}\
S_{vN}=\sum_j\Big{[}(\nu_j+\frac{1}{2})\ln (\nu_j+\frac{1}{2})-(\nu_j-\frac{1}{2})\ln(\nu_j-\frac{1}{2})\Big{]},\nonumber
\end{equation*}
where $\nu_j$'s are the eigenvalues of the matrix $(\mathbf{X}_A\mathbf{P}_A)^{\frac{1}{2}}$, where $\mathbf{X}_A=\frac{1}{2}(\mathbf{K})^{-\frac{1}{2}}|_A$
and $\mathbf{P}_A=\frac{1}{2}(\mathbf{K})^{\frac{1}{2}}|_A$ are the two point correlations of the position and the momentum restricted to the subsystem $A$.

\setcounter{equation}{0}
\renewcommand{\theequation}{C\arabic{equation}}

\subsection*{Appendix C: Structure of degeneracies and average entanglement over excited states of the XX-chain}\label{app3}

In this appendix, we tend to present a detailed study of various averagings of entanglement entropy for a periodic 
free fermion system in excited states. By excited states, we mean eigenstates of Hamiltonian produced by the action
of $\eta^{\dagger}_k$ on the vacuum state of the XX-chain; Among these states, there are sets of degenerate states. 
In \cite{Vidmar2017}, the authors calculated the entanglement entropy for all the states produced as explained above, and then they took
the average of all the entropies calculated without counting for degenerate states. Here, we revisit the work of \cite{Vidmar2017} by taking 
into account the degeneracies. First, we identify the degenerate states and then preform various averagings on the entanglement entropies.

\noindent\textbf{Degenerate Energy States:}
The XX-chain can be solved exactly by diagonalizing the Hamiltonian using the Fourier transformation, as written in the main text. 
Each excited state can be produced by acting on the vacuum with a set of creation operators with different modes. Each mode $\eta_k$ has energy
\begin{equation}\label{eq:Sup modeenergy}
	|f(k)|; \qquad k=1,2,\cdots,L,
\end{equation}
where $L$ is the size of the system. For instance, states $|\psi^m\rangle$ and $|{\psi^n}\rangle$ can be degenerate ($E^m=E^n$) but they are made of different combinations of $\eta^\dagger$'s. 
\begin{equation*}
	\begin{aligned}
        |{\psi^m}\rangle&=\eta^\dagger_{m_1}\eta^\dagger_{m_2}\eta^\dagger_{m_3}\cdots|{0}\rangle,\\
        |{\psi^n}\rangle\:&=\eta^\dagger_{n_1}\eta^\dagger_{n_2}\eta^\dagger_{n_3}\:\cdots\:|{0}\rangle,
		\end{aligned}
\end{equation*}
in a way that
$$\{k_{m_1},k_{m_2},k_{m_3},\cdots\}\neq\{k_{n_1},k_{n_2},k_{n_3},\cdots\}.$$

\begin{figure}[t]
    \centering
    \includegraphics[width=0.5\textwidth]{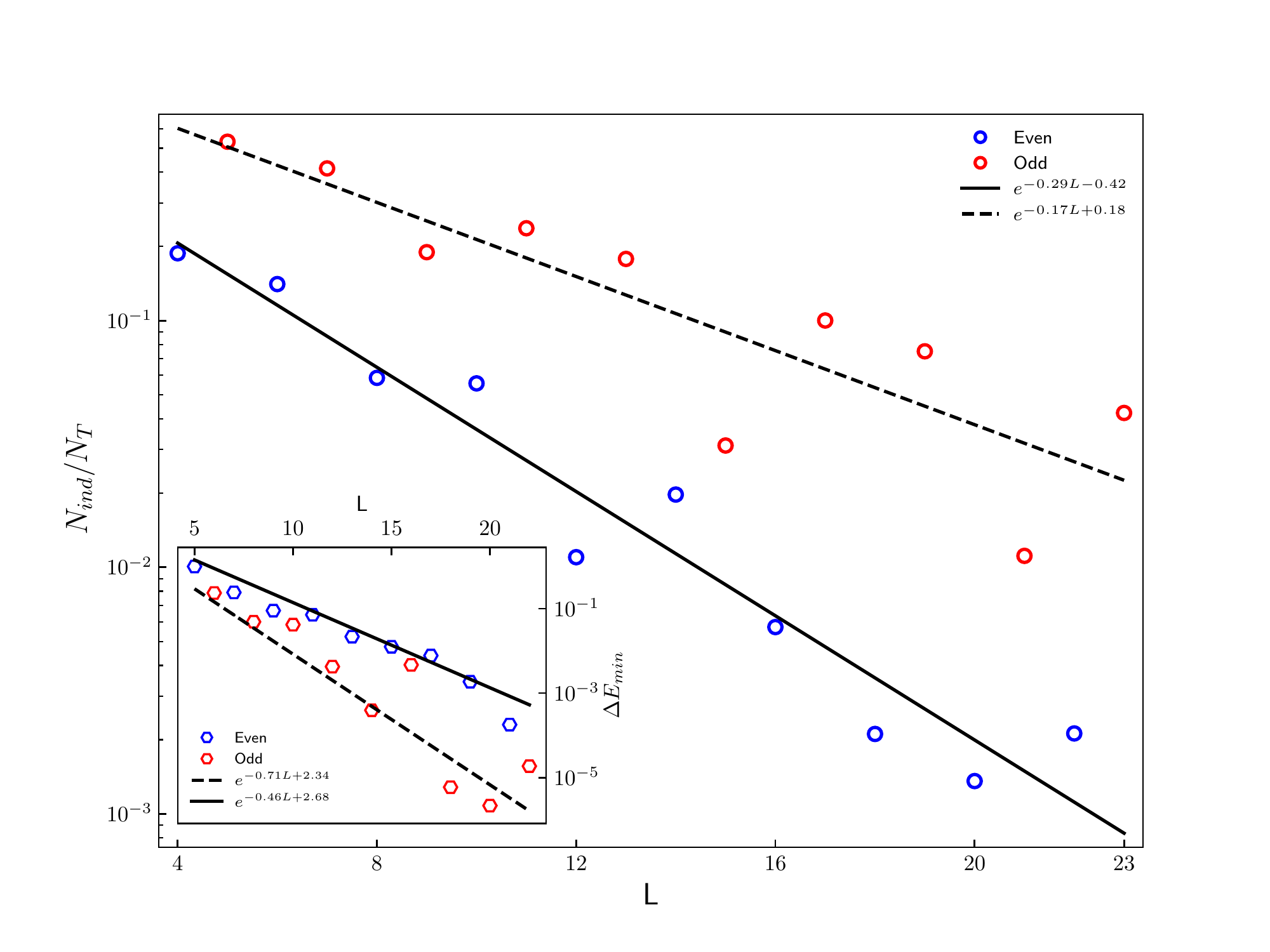}
	\caption{\footnotesize In this figure, the ratio of independent energies to total number of energies is plotted versus the system size. $\frac{N_{ind}}{N_T}$ decreases exponentially with size of the system. (Inset) Plot of $\Delta E_{min}$ versus the size of the system. This quantity gives us a measure to set a precision for finding the unequal energy levels of the spectrum. Note that for $L$ even we have degeneracy even in the level of the energy modes which is the main reason for stronger decay of $\frac{N_{ind}}{N_{T}}$ with respect to the $L$ odd case.}\label{fig:Sup IndEnergies}
\end{figure}

For instance in $XX$ model, the energy of the modes with $k=l$ and $k=L-l$ is the same, 
therefore the degeneracy in the eigenstates is expected. To unravel these groups of degenerate eigenstates as accurate as possible (machine precision), 
we first calculated the minimum energy gap ($\Delta E_{min}$) of the spectrum. Since in these types of Hamiltonians the energy levels are not equally 
spaced, the $\Delta E_{min}$ helps us to have an idea for the required precision value to decide whether two energies are equal or not. Not surprisingly, as shown in the inset of 
the figure \ref{fig:Sup IndEnergies}, this gap decreases exponentially with the size of the system which makes the decision that two states are degenerate or not more difficult by increasing the size of the system.
After finding all the degeneracies, we sum over all the non-degenerate states and call the number $N_{ind}$. This number which we loosely call the number of independent states
is exponentially smaller than the dimension of the Hilbert space. In other words, as shown in the figure \ref{fig:Sup IndEnergies}, the ratio $\frac{N_{ind}}{N_T}$ decays exponentially 
with the size of the system $L$. This indicates the presence of an enormous amount of degeneracies in the spectrum of the XX-chain.

\noindent\textbf{Averaging Types:} Because of the huge degeneracy in the spectrum of periodic free fermions it is clear that one has a lot of freedom in choosing an eigenstate with different amount of entanglement.
The more degeneracy means that there is more freedom in picking a state with very large or very small entanglement content. The sum of two maximmaly entangled states can be in principle a product state
with zero entanglement. The reverse is also true, the sum of two product states can be maximmaly entangled. When there is a degeneracy in the spectrum
it is not feasible to look for a state with max/min entanglement. We take another approach by focusing on the eigenstates that one can find by the method of previous section.
After creating all the eigenstates, we calculated the entanglement entropy using the Peschel formula \cite{peschel2003calculation} which is valid also for the excited eigenstates.
Note that this is not the case for an arbitrary state due to the lack of Wick's theorem.
After getting the entanglement entropy for all the eigenstates, we averaged these entanglements for $L_A$ from $1$ to $L-1$ in four different ways.
\begin{itemize}
	\item Averaging over all the entanglement entropies without counting for degeneracies (VHBR mean), as in \cite{Vidmar2017}.
	\item Taking the average of entropies of degenerate eigenstates, and then calculating the average of resulting entropies (WA mean).
	\item First, finding the minimum value of the entropy in any set of the degenerate eigenstates, and then calculating the average of this minimum entropies (MIN mean).
	\item Finding the maximum value of entropy in any set of degenerate eigenstates, and then calculating the average of this maximum entropies (MAX mean).
	\end{itemize}

\bibliography{ExtEntRef}

\end{document}